\documentclass[12pt]{article}

\usepackage[final]{graphicx}
\usepackage{amsmath}
\usepackage{amssymb}
\usepackage{xcolor}
\usepackage{tensor}
\usepackage{braket}
\usepackage{hyperref}
\usepackage{units}
\usepackage{MnSymbol}

\newcommand{\tffH}{{\mathcal H}}
\newcommand{\tffE}{{\mathcal E}}
\newcommand{\tffF}{{\mathcal F}}

\newcommand{\SigG}{{\text{S}}}

\newcommand{\Q}{{\cal Q}}

\newcommand{\xB}{x_{\rm B}}

\newcommand{\g}{{\text{G}}}

\newcommand{\GeV}{{\rm GeV}}

\def\parameters{\mbox{\boldmath $p$}}

\def\mQ2{\mathcal Q^2}

\usepackage{ifthen}
\newcommand{\prob}[2][\empty]{
  \ifthenelse{\equal{#1}{\empty}}
    {P\! \left( #2  \right) }
    {P\! \left(  #1  \middle\vert #2 \right)}
}
\newcommand{\probe}[2][\empty]{
  \ifthenelse{\equal{#1}{\empty}}
    {P\! \left[ #2  \right] }
    {P\! \left[  #1  \middle\vert #2 \right]}
}

\newcommand{\cprob}[2][\empty]{
  \ifthenelse{\equal{#1}{\empty}}
    {f\! \left( #2  \right) }
    {f\! \left(  #1  \middle\vert #2 \right)}
}

\newcommand{\cond}[2][\empty]{
  \ifthenelse{\equal{#1}{\empty}}
    {\left( #2  \right) }
    {\left(  #1  \middle\vert #2 \right)}
}

\usepackage{multirow}
\usepackage{suffix}
\usepackage{subfigure}
\usepackage{placeins}
\usepackage{amsmath}
\usepackage{environ}

\usepackage{amssymb}
\newcounter{comment}

{\refstepcounter{comment}%
\begin{quote}
\ttfamily\small$\blacksquare$ \textbf{\underline{Comment} $\sharp$\thecomment:}}%
{\end{quote}}

{
\begin{quote}
\ttfamily\small$\blacktriangleright$ \textbf{\underline{Reply} $\sharp$\thecomment:}}%
{\end{quote}}

\def\muFgpd{\relax\ifmmode\mu_\text{F,GPD}^2\else{$\mu_\text{F,GPD}^2${ }}\fi}
\def\muFda{\relax\ifmmode\mu_\text{F,DA}^2\else{$\mu_\text{F,DA}^2${ }}\fi}
\def\muO{\relax\ifmmode{\mu_{0}^{2}}\else{$\mu_{0}^{2}${ }}\fi}
\def\Mev{\relax\ifmmode{\text{MeV}}\else{MeV{ }}\fi}

\def\Li{\relax\ifmmode{\text{Li}_{2}}\else{Li$_2${ }}\fi}

\newcommand{\fTFF}[1]{\tensor[^a]{\mathcal F}{^A}}

\usepackage{ifthen}

\newcommand{\lh}[2][\empty]{
  \ifthenelse{\equal{#1}{\empty}}
    {L \left( #2  \right) }
    {L \left( \left. #1  \right| #2 \right)}
}

\font\cmss=cmss12 
\def\1{\hbox{{1}\kern-.25em\hbox{l}}}
\def\bfZ{\relax{\hbox{\cmss Z\kern-.4em Z}}}
\usepackage[colorinlistoftodos]{todonotes}

\begin{document}

\begin{titlepage}

{\ }\\

\centerline{\large \bf Global analysis of generalized parton distributions}
\centerline{\large \bf -- collider kinematics --}

\vspace{15mm}

\centerline{T.~Lautenschlager$^a$, D. M\"uller$^b$ and A.~Sch\"afer$^a$}

\vspace{8mm}

\centerline{\it $^a$ Institut f\"ur Theoretische Physik, University Regensburg}
\centerline{\it D-93040 Regensburg, Germany}
\vspace{4mm}

\centerline{\it $^b$Institut f\"ur Theoretische
Physik II, Ruhr-Universit\"at Bochum}
\centerline{\it  D-44780 Bochum, Germany}

\vspace{25mm}

\centerline{\bf Abstract}

\vspace{5mm}

\noindent
We utilize H1 and ZEUS data for exclusive electroproduction of photons, $\rho^0$- and $\phi$-mesons to access generalized parton distributions at small momentum fraction. To do so, we employ state-of-the art techniques, based on next-to-leading order perturbation theory, flexible model parametrization, and Bayesian inference. We provide a partonic interpretation of our analysis, where emphasize is given to the transverse distribution of sea quarks and gluons.

\vspace{0.5cm}

\noindent

\vspace*{52mm}
\noindent
Keywords: hard exclusive electroproduction, vector mesons, generalized parton distributions

\noindent
PACS numbers: 12.38.Bx 12.39.St 13.60.Fz 13.60.Le

\end{titlepage}


\section{Introduction}

Generalized parton distributions (GPDs) \cite{Mueller:1998fv,Radyushkin:1996nd,Ji:1996nm} allow to quantify the partonic structure, in particular of the nucleon, from a comprehensive perspective. One important aspect of this framework is the possibility to access the transverse distribution of partons in a perturbative approach where unobservable degrees of freedom are mostly integrated out. Due to strong arguments for the validity of factorization, deeply virtual Compton scattering (DVCS) is considered as a theoretically clean process to constrain GPDs. However, this is a challenging task since an unbiased analysis requires a complete measurement of at least twenty-four  azimuthal angular harmonics to deduce twelve complex valued Compton form factors (CFFs) for a given threefold kinematical bin, the latter being described by the values of the Bjorken scaling variable $\xB$, the $t$-channel momentum transfer squared $t$, and the photon virtuality squared $-\Q^2$ \cite{Belitsky:2001ns,Kumericki:2013br}. Note that in DVCS kinematics several of these CFFs are power, perturbatively, or kinematically suppressed; such that they can be set to zero to facilitate the task. Deeply virtual meson electroproduction (DVMP) where the virtual photon is longitudinally polarized offers further possibilities to access GPDs, in particular, these processes allow flavor separation. Here a complete access to the transition form factors (TFFs) requires also a polarization transfer measurement \cite{Mueller:2013caa}. Four photon helicity non-flip CFFs and two longitudinally polarized photon TFFs factorize in terms of hard amplitudes and twist-two GPDs
\cite{Collins:1998be,Collins:1996fb}. The hard amplitudes and evolution kernels are perturbatively known to next-to-leading order (NLO) accuracy. While for DVCS a global GPD analysis was successfully developed \cite{Kumericki:2007sa,Kumericki:2009uq,Kumericki:2010fr,Kumericki:2013br}, the DVMP processes were not systematically treated in the perturbative framework.  Apart from the hand bag approach \cite{Goloskokov:2005sd,Goloskokov:2007nt,Goloskokov:2009ia} only little effort was devoted to $\pi^+$ and vector meson production (in small-$\xB$ kinematics), namely to leading order (LO) accuracy in \cite{Bechler:2009me,Meskauskas:2011aa}.
From these findings and information on GPDs, derived from DVCS data in LO approximation \cite{Meskauskas:2011aa,Kumericki:2011zc,Kroll:2012sm}, one expects that a global GPD analysis at LO is hardly possible, whereas a NLO analysis should be more consistent.

In the following we study the access to GPDs in the small-$\xB$ region, where DVCS, DV$\!\rho^0_{\rm L}$P, and DV$\!\phi_{\rm L}$P cross sections were measured by the HERA collider experiments H1 and ZEUS. Contrarily, to the common view that the onset
of the perturbative regime might happen at very large scales, like $\Q^2 \sim 15\, \GeV^2$, e.g., see \cite{Frankfurt:2005mc}, the  DV$V^0_{\rm L}$P data might be perturbatively describable for $\Q^2 \gtrsim 4\, \GeV^2$, see \cite{Meskauskas:2011aa} for a discussion. The $t$-channel exchange of the `pomeron', viewed as a gluon- and quark-pair exchange in the color singlet state, ensures that the cross sections do not vanish in the high-energy limit.
Since the `pomeron'  couples in the first place to the proton helicity conserving CFF ${\cal H}$ or TFF ${\cal H}_{V^0}$ the phenomenology  simplifies drastically in the small-$\xB$ region. For instance, the $t$-differential DVCS  and DVMP cross sections at small-$\xB$ can be safely approximated as
\begin{eqnarray}
\label{dsigma}
\frac{d\sigma^{\gamma_{\rm T}^*\,p\to \gamma\,p}}{dt} \approx
%
\frac{\pi \alpha_{\rm em}}{\Q^4}
\left|\xB \tffH(\xB,t,\Q^2)\right|^2 
\quad\mbox{and}\quad
\frac{d\sigma^{\gamma_{\rm L}^*\,p\to V_{\rm L}^0\,p}}{dt} \approx
  \frac{(2\pi)^2 \alpha_{\rm em}}{\Q^4}
\left|\xB \tffH_{V_{\rm L}^0}(\xB,t,\Q^2)\right|^2
,
\end{eqnarray}
where $\alpha_{\rm em}\approx 1/137$ is the electromagnetic fine structure constant
and the Hand convention
was utilized to convert the electroproduction cross sections into the virtual photon ones.  Here, we neglected
$|\tffE|^2$ and  $|\tffE_{V^0}|^2$ terms, although
it can not be excluded that the `pomeron' couples to the proton helicity flip CFF/TFF $\tffE$/$\tffE_{V^0}$. However, these contributions
are kinematically suppressed by a pre-factor $t/4 M_N^2$ and since the average $\langle t\rangle$ is a small quantity compared to the nucleon mass
square $M^2_N$, it is justified to neglect the proton helicity flip contributions. Furthermore, in the group  theoretical
decomposition of CFFs and TFFs the flavor non-singlet part is less important and so they are mainly governed by the singlet (S) contributions,
e.g., for four active quarks and a isospin symmetric quark sea we express the TFFs in terms of $\mathrm{SU}(n_f=4)$ multiplets:
\begin{subequations}
\label{cfftffF-flavor}
\begin{eqnarray}
\label{cffF-flavor}
\tffF &\!\!\!\approx\!\!\!& \frac{1}{18}\tffF^{8^{(+)}}-\frac{1}{18}\tffF^{15^{(+)}}+\frac{5}{18} \tffF^{\SigG}\,,
\\
\label{tffF-flavor}
\tffF_{\rho^0} &\!\!\!\approx\!\!\!&
\frac{1}{6\sqrt{2}}\tffF_{\rho^0}^{8^{(+)}}+ \frac{1}{12\sqrt{2}}\tffF_{\rho^0}^{15^{(+)}}+
\frac{1}{\sqrt{2}} \tffF_{\rho^0}^{\SigG}
\,,\quad
\tffF_{\phi}  =
\frac{1}{9} \tffF_{\phi}^{8^{(+)}}- \frac{1}{36}  \tffF_{\phi}^{15^{(+)}}- \frac{1}{3} \tffF_{\phi}^{\SigG}\,.
\end{eqnarray}
\end{subequations}
Note also that at small-$\xB$ the real and imaginary parts are (approximately) related to each other.

\section{Model and data specifications}

The CFFs [TFFs] are evaluated from a Mellin-Barnes integral, see \cite{Kumericki:2007sa,Kumericki:2009uq,Mueller:2013caa} and references therein, depending on the conformal GPD [and DA] moments $ H^p_j(\xi,t,\Q_0^2)$ with $p\in \{{\rm NS}^{(+)},\Sigma,{\rm G}\}$ [and $\varphi_{k\, {\rm  V}^0_L}(\Q_0^2)$] that are specified at the input scale $\Q_0=2\, \GeV$. The flavor singlet TFF reads
\begin{eqnarray}
\label{tffH_{V_L^0}}
\tffH_{V_L^0}^{\SigG}(\xB,t,\Q^2) &\!\!\! \stackrel{\rm Tw-2}{=}\!\!\!  &
\frac{4 f_{{\rm  V}^0_L}}{9 \Q}  \sum_{{k=0\atop {\rm even}}}\frac{1}{2 i} \int_{c- i \infty}^{c+ i \infty}\! dj\, \xi^{-j-1}
\left[
  i + \tan\!\left(\!\frac{\pi\,j}{2}\!\right)
  \right]
\\
&&\phantom{\frac{4 f_{{\rm  V}^0_L}}{9 \Q}  \sum_{{k=0\atop {\rm even}}}\frac{1}{2 i}\int_{c- i \infty}^{c+ i \infty}\! dj\,}
\times \varphi_{k\,{\rm  V}^0_L}(\Q_0^2)\,
   \mbox{\bf T}_{jk}\!\left(\!\xi\big|\alpha_s(\Q),\alpha_s(\Q_0)\right)\cdot
\left({ H^\Sigma_j(\xi,t,\Q_0^2) \atop H^\g_j(\xi,t,\Q_0^2) }\right)
\,,
\nonumber
\end{eqnarray}
where $\xi=\xB/(2-\xB)$,  the vector valued hard amplitude $\mbox{\bf T}_{jk}$, containing the evolution operators, is defined for the $\overline{\rm MS}$-scheme in \cite{Mueller:2013caa}, and the meson decay constants are $f_{\rho^0} = 209\, {\rm MeV}$ and  $f_{\phi} = 221\, {\rm MeV}$.
Analogous formulae hold true for the flavor non-singlet sector (one simply has to replace vector valued quantities by non-singlet ones). Those  for the CFF $\cal H$  can be found in \cite{Kumericki:2007sa,Kumericki:2009uq}. Note that our framework is related to the `dual' parametrization as given in \cite{Polyakov:2007rw}.  The $\overline{\rm MS}$-scheme value of the strong coupling, we use, is
$\alpha_S^{\text{NLO}}(\Q_0=2\, \GeV) = 0.291$.
The normalization of the TFF depends also on the meson DA, which is essentially unknown from phenomenology, however, its first (conformal) moments can be calculated from QCD sum-rules \cite{Ball:1998sk,Pimikov:2013usa} and they turn out to be small.

Our flexible model for the sea quark and gluon GPDs $H$ is an extension of the one in \cite{Kumericki:2007sa,Kumericki:2009uq}. As there, we do not incorporate the result of a global PDF estimation, rather we include a limited data set on the DIS structure function $F_2$ in our analysis%
\footnote{This is caused by the fact that process independent PDFs are conventionally defined. This is rather uncritical at LO accuracy, however, in the small-$\xB$ region the rather sizeable radiative NLO corrections are also sensitive to the scheme conventions. Hence, for a standard PDF parametrization one has to take the same partonic specifications as used for their inference. This may require some additional theoretical work, postponed to the future, in particular for the treatment of charm quarks \cite{Noritzsch:2003un}, which we consider for simplicity as a forth light quark here.}.
The PDF models, already decorated with $t$-dependencies, are given at the input scale $\Q_0$ and the (standard) Mellin moments read
\begin{subequations}
\label{H_j}
\begin{eqnarray}
\label{H_j^Sigma}
H_j^\Sigma(t,\Q_0^2|\alpha^{\rm sea},\beta^{\rm sea},M^{\rm sea}) &\!\!\! =\!\!\!& 
N^{\rm sea}\, \frac{\left[j+1-\alpha^{\rm sea}\right]B(1-\alpha^{\rm sea} +j,\beta^{\rm sea}+1)}{
\left[j+1-\alpha^{\rm sea}(t)\right]B(2-\alpha^{\rm sea},\beta^{\rm sea}+1)}\, \frac{1}{\left(1-t/M^{{\rm sea}\,2}\right)^2} +\cdots,
\nonumber\\
\\
\label{H_j^g}
H_j^\g(t,\Q_0^2|\alpha^{\g},\beta^{\g},M^{\g}) &\!\!\! =\!\!\!& 
N^{\rm G}
\frac{\left[j+1-\alpha^{\g}\right]B(1-\alpha^{\g} +j,\beta^{\g}+1)}{\left[j+1-\alpha^{\g}(t)\right]B(2-\alpha^{\g},\beta^{\g}+1)} \frac{1}{\left(1-t/M^{{\g}\,2}\right)^2} \,,
\end{eqnarray}
\end{subequations}
where the ellipsis stand for valence quark contributions that can be neglected at small-$\xB$.
The large $j$- (or $x$) behavior is rather uncritical for our purpose and it is fixed by $\beta^{\rm sea}=8$ and $\beta^{\g}=6$,
see discussion in \cite{Kumericki:2009uq}.
The standard Mellin moments of unpolarized quark singlet
and gluon PDFs,
$$\Sigma_j(\Q_0^2) = H_j^\Sigma(t=0,\Q_0^2|\cdots)\quad\mbox{and}\quad g_j(\Q_0^2) = H_j^\g(t=0,\Q_0^2|\cdots)\,,$$
are normalized for $j=1$ to the momentum fraction $N^{\rm sea}$ and $N^{\g}$, respectively. We utilize the sum rule
$N^{\rm sea} + N^{\rm val} + N^{\rm G} =1$, where in agreement with common PDF parameterizations
we set $N^{\rm val}=0.4$.
The three remaining PDF parameters, i.e., the `pomeron' intercepts $\alpha^{\rm sea}$ and $\alpha^{\g}$ as well as the momentum fraction $N^{\rm sea}$, are sufficient for a good  description of DIS data \cite{Kumericki:2007sa,Kumericki:2009uq}.

The position of the `pomeron' pole at $j+1=\alpha^p(t)$ is in our models (\ref{H_j}) given by the trajectory $\alpha^p(t) = \alpha^p + \alpha^{\prime\, p}\,t $ with $p\in\{{\rm sea}, \g\}$. The residual $t$-dependencies are parameterized by dipoles with cut-off masses $M^p$. Note that these masses are expected to vanish at large $j$, however, in the small-$x$ region this feature can not be addressed and, thus, we do not decorate them with an index $j$.
Our GPD models are formulated in terms of (conformal) moments, depending now on the skewness parameter $\eta$, and are expanded in terms of three {\em effective} SO(3)-PWs
\begin{eqnarray}
\label{H^p_n}
  H^p_n(\eta,t,\Q^2_0) = \sum\limits_{{J =n-3 \atop {\rm even}}}^{n+1} s_{n+1-J}\, H^p_{J-1}(t,\Q^2_0|\alpha_{n+1-J}^p,M_{n+1-J}^{p}) \; \eta^{n+1-J} \; \hat d_{\cal H}^J(\eta) \quad \mbox{for} \quad n\in \{ 3, 5, 7, \cdots\}\,.
\end{eqnarray}
Here $d_{\cal H}^J(\eta)$ is a $t$-channel SO(3)-PW which is normalized at  $\eta=0$ to one, $\nu=n+1-J$ labels the SO(3)-PWs ($\nu=0$ leading (l), $\nu=2$ next-leading (nl), and so on), the PW amplitudes $ H^p_{J-1}$ are defined in (\ref{H_j}), and the strength of non-leading ones is controlled by the parameters
$s_2^p$ and $s_4^p$.  In our model  the `pomeron' behavior is implemented as poles at $J=\alpha^p(t)$, i.e., it is associated with $t$-channel angular momentum $J$ rather than with the conformal spin $j+1$. This provides us with the possibility to control the residues and even their evolution flow with the strength parameters $s_2^p$ and $s_4^p$ %
\footnote{If only the leading SO(3)-PW contributes, one arrives at the claim that certain GPDs at small-$x$ and $t=0$ can be obtained from PDFs \cite{Shuvaev:1999ce,Martin2009zzb}.}.
For reasons spelled out below, we extended previous GPD parameterizations by allowing each SO(3)-PW to have its own parameter set $\{\alpha_{\nu}^p,M_{\nu}^p\}$, which gives a more flexible parametrization.
Finally, the ansatz (\ref{H^p_n}) is plugged into the Mellin-Barnes integrals (\ref{tffH_{V_L^0}}),
where we perform in non-leading SO(3)-PW amplitudes a shift of the integration variable $j$ \cite{Kumericki:2009uq}, and predicts  the
cross sections (\ref{dsigma}). Our statistical data analysis allows then to extract the GPD  parameters.

For us it is sufficient to use the DIS data from the H1 collaboration \cite{Aid:1996au}
(we did not make an attempt to describe the combined H1 and ZEUS data set \cite{Aaron:2009aa}).
The DVCS cross section%
\footnote{It was obtained from the unpolarized electroproduction cross section by subtracting the Bethe-Heitler one, where azimuthal angular integration suppresses the interference term.}
was measured by the H1 \cite{Adloff:2001cn,Aktas:2005ty,Aaron:2007ab,Aaron:2009ac} and ZEUS \cite{Chekanov:2003ya,Chekanov:2008vy} collaborations and
it is considered to be dominated by the (transversal) photon helicity conserved DVCS amplitude square.
We use only the data from \cite{Aktas:2005ty,Aaron:2009ac,Chekanov:2008vy} that can be considered as statistical independent.
In DV$V^0$P the H1 and ZEUS collaborations extracted primarily the photoproduction cross section that stems from transversally and longitudinally polarized photons
\begin{align}
  d\sigma(t|W,\Q^2) = d\sigma_\text{T}(t|W,\Q^2) + \varepsilon(W,\Q^2) \; d\sigma_\text{L}(t|W,\Q^2)\,,
  \quad
\varepsilon\left(W,\Q^2\right) \approx \frac{1-y}{1-y+\frac{1}{2}y^2}\,,
\end{align}
where  the flux ratio $\varepsilon(y)$ of longitudinal to transverse photons is depending on the electron energy loss $y = (W^2 + \Q^2 -M^2)/(s - M^2)$ which is a function of $W$, $\Q^2$ and the center-of-mass energy $s$ that is fixed.  In the following we take the measurements for the $\rho^0$ and $\phi$ channels \cite{Chekanov:2005cqa,Chekanov:2007zr,Aaron:2009xp} and ignore the statistically less important data from the $\omega$ channel \cite{Breitweg:2000mu}.
Relying on the hypothesis of $s$-channel helicity conservation, the H1 and ZEUS collaboration were also able to measure the
ratio of the longitudinal  and transverse cross section
\begin{align}
\label{R(cQ^2)}
  R(\Q^2) = \frac{ \sigma_\text{L}(\langle W \rangle, \Q^2) }{ \sigma_\text{T}(\langle W \rangle,\Q^2) }\,,
\end{align}
as function of $\Q^2$ for the mean $\langle W \rangle \sim 75\, \GeV$ and integrated over $0 \lesssim -t \lesssim 0.5\, \GeV^2$. To access twist-two GPDs
we need only the longitudinal cross section. To include the complete set of data, we employ
\begin{align}
  \label{Eq: sigmaTot to sigmaL}
  \frac{d\sigma_\text{L}}{dt} \left( W,t, \Q^2\right) = \frac{1}{R^{-1}\left(\Q^2\right) +\varepsilon(W,\Q^2) } \,  \frac{d\sigma \left(W,t, \Q^2\right)}{dt}\,.
\end{align}
The $R$-ratio, vanishing for real photoproduction, is modeled as in \cite{Meskauskas:2011aa}
\begin{align}
\label{R-model}
R(\Q^2) = \frac{\Q^2}{m_{\rm V}^2} \left( 1 + a \frac{\Q^2}{m_{\rm V}^2} \right)^{-p} \quad\mbox{with}\quad  m_\rho = 0.776\, \GeV
\mbox{ and }
m_\phi = 1.02\, \GeV\,,
\end{align}
where a possible $W$- and $t$-dependence, which can not be resolved by the existing experimental data, is neglected.
Note that the power in $\Q^2$ predicted by dimensional counting for the large-$\Q^2$ asymptotics, i.e., $p=0$, is modified to be $\Q^{2-2p}$.


\section{Data analysis utilizing Bayesian inference}

To outline the principles of our analysis, we present a brief introduction to Bayesian inference. According to the sum and product rule of probability theory \cite{Jaynes2003}, the  {\it posterior} probability distribution function (pdf) of a set of parameters $\mbox{\boldmath $p$}$ reads
\begin{align}
  \label{Eq: Bayes Theorem}
  \cprob[\parameters]{D I} = \frac{ \cprob[\parameters]{I} \prob[D]{\parameters I} }{ \prob[D]{I} } \, ,
\end{align}
where $D$ symbolizes the given data and $I$ stands for all available prior knowledge besides the data. The individual terms are the {\it prior} pdf $\cprob[\parameters]{I}$ for the parameters, the normalization constant $\prob[D]{I}$ and the likelihood $\prob[D]{\parameters I}$. Note that the symbol $\parameters$ stands for the logical product of the parameters. In contrast, the well known method of least squares  maximizes the likelihood with respect to the parameters and neglects the prior pdf and the normalization constant\footnote{In case of a linear model and a prior which is dominated by the data, the least squares estimate is a special case of Bayesian inference. However, these requirements are not fulfilled in our problem and the error estimation via the covariance matrix will be questionable.}. The knowledge of the normalized posterior pdf for the parameters is equivalent to the solution of the parameter estimation problem. For example, the pdf for a quantity $q$ given as a function $q=Q(\mbox{\boldmath $p$})$ of the parameters is %
\begin{align}
  \label{eq: Error Propagation}
  \cprob[q]{DI} = \int\!\! d\parameters\, \delta(q-Q(\mbox{\boldmath $p$})) \cprob[\parameters]{D I}\,.
\end{align}
The mean value and the variance of the single parameter $p_i$,
\begin{align}
  \mathrm E(p_i) = \int\!\! d\parameters\, p_i \; \cprob[\parameters]{DI} \quad\mbox{and}\quad  \mathrm{Var}(p_i) = \mathrm E(p_i^2) - \mathrm E^2(p_i)\,,
\end{align}
respectively, are (central) moments of the posterior pdf.

\begin{figure}[t]
  \centering
  \includegraphics{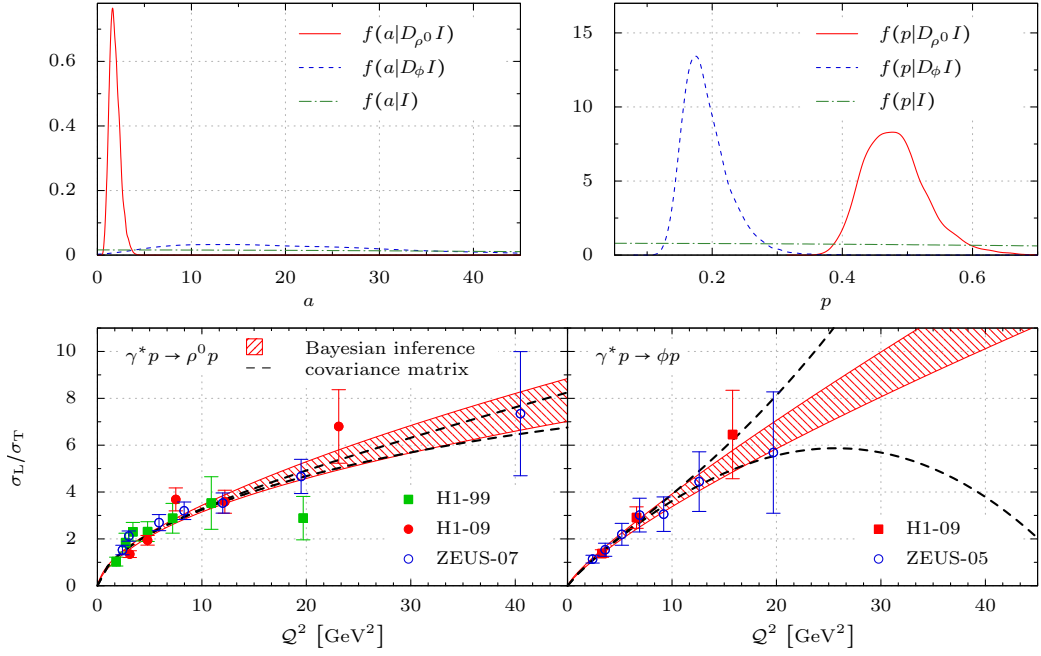}
\vspace{-3mm}
  \caption{\small Prior (dash-dotted) and posterior (solid, dashed) pdfs of the $R$-ratio model (\ref{R-model}) parameters $a$ (top, left) and $p$ (top, right). Experimental data of the $R$-ratio (\ref{R(cQ^2)}) from \cite{Adloff:1999kg,Aaron:2009xp,Chekanov:2007zr} and \cite{Aaron:2009xp,Chekanov:2005cqa} for DV$\!\rho^0$P (bottom, left) and DV$\!\phi$P (bottom, right). The hatched area corresponds to the square root of the variance of the ratio, while the dashed lines show the same quantity from a conventional least square fit.}
  \label{fig:pdf of the ratio parameters a and p}
\end{figure}

Applying this approach, the inference of the parameters $a$ and $p$ for the $R$-ratio (\ref{R(cQ^2)}) provides the posterior pdf $\cprob[a p]{DI}$. In the upper row in Fig.~\ref{fig:pdf of the ratio parameters a and p} we show the pdfs for both parameters (solid and dashed curves for DV$\!\rho^0$P and DV$\!\phi$P, respectively), which we obtain by marginalization together with their prior distributions (flat dash-dotted curves). The parameters for the $R$-ratio of DV$\!\rho^0$P are well constrained by the data, thus the posterior pdf surmounts the prior distribution and has a smaller variance. This is also true for the parameter $p$ for the $R$-ratio of DV$\!\phi$P. However, we realize that due to the missing data above $\Q^2 \sim \unit[20]{GeV^2}$ the posterior pdf for the parameter $a$ from DV$\!\phi$P is restricted by the prior and can not be determined accurately.
To report the posterior pdfs for the parameters, we present the mean values and the square root of the variances:
\begin{align}
a_{\rho^0} = 1.845 \pm 0.554,\; p_{\rho^0} = 0.482 \pm 0.0481,\quad
a_\phi = 22.181 \pm 13.450 ,\; p_\phi = 0.191 \pm 0.035\,.
\end{align}
Our estimate (hatched error band) is displayed together with the one from a conventional least square fit (dashed lines surround the error band obtained via the covariance matrix) and experimental measurements in the two lower panels of Fig.~\ref{fig:pdf of the ratio parameters a and p} for DV$\!\rho^0$P (left) and DV$\!\phi$P (right).
Employing the $R$-ratio, we are able to extract the longitudinal cross section from the total one (\ref{Eq: sigmaTot to sigmaL}). For a fixed kinematic point, the pdf for the $R$-ratio $r=R\left(\mathcal Q^2\right)$ is given by (\ref{eq: Error Propagation}):
\begin{align}
  \cprob[r]{D I} = \int\!\!\!\! \int\!\!dadp \; \delta\left( r - R\left(\mathcal Q^2\right)  \right) \cprob[ap]{D I} \,.
\end{align}
Hence, we derive the pdf for the longitudinal cross section from the data $D^\prime$ for the total one by
\begin{align}
  \cprob[\sigma_\text{L}]{D D^\prime I}  =  \int\!\! d\sigma\!\!\int\!\!dr \;
   \delta\!\left(\! \sigma_L - \frac{1}{r^{-1}+\varepsilon} \sigma \!\right) \; \cprob[r]{DI} \cprob[\sigma]{D^\prime I}\,.
\end{align}

In our analyses of the statistical independent DIS \cite{Aid:1996au}, DVCS \cite{Aktas:2005ty,Aaron:2009ac,Chekanov:2008vy}, DV$\!\rho^0$P
\cite{Chekanov:2007zr,Aaron:2009xp}, and DV$\!\phi$P \cite{Chekanov:2005cqa,Aaron:2009xp}  measurements we include all data above $\Q^2 = 4\, \GeV^2$.
The normalization uncertainties of these data are taken into account as additional parameters ${\cal N}_{\rm data}$, where for DVCS these uncertainties are
not given (${\cal N}_{\rm DVCS}\equiv 1$). Our goal is to estimate the $2\times 10$ GPD parameters,
\begin{align}
\label{ps}
\mbox{\boldmath $p$}=\mbox{\boldmath $p$}^{\rm sea} \cup \mbox{\boldmath $p$}^{\rm G} \quad\mbox{with}\quad
\mbox{\boldmath $p$}\indices{^{p}}=\{N\indices*{^{p}},\alpha^{p},\alpha\indices*{^{\prime\,p}},M\indices*{^{p}},s\indices*{_2^{p}},\alpha\indices*{_2^{p}},M\indices*{_2^{p}},s\indices*{_4^{p}},\alpha\indices*{_4^{p}},M\indices*{_4^{p}}\}
\quad\mbox{for}\quad p \in \{{\rm sea}, {\rm G}\},
\end{align}
of our sea quark and gluon models $H_j^p(\eta,t,\Q_0^2|\mbox{\boldmath $p$}^p)$ in (\ref{H^p_n}) from data set $D$ under the assumptions $I$, we spelled out above. In addition, we use prior pdfs to constrain the parameters. A narrow Gaussian distribution incorporates the sum rule $N^\text{sea}+N^\text{G}+N^\text{val}=1$ with $N^\text{val}=0.4\pm 0.05$. In this way, we implement the sum rule, but also test the assumption of the normalization parameter of the valence contribution. Using the same intercept and cut-off mass parameters for all SO(3)-PWs was very successful to describe the data for DIS and DVCS in LO, NLO and NNLO \cite{Kumericki:2007sa,Kumericki:2009uq}. For the inclusion of DV$V^0$P, the challenging task at LO is to describe DIS data \cite{Meskauskas:2011aa}. Thus, we additionally introduce new $\alpha_\nu^p$ and $M_\nu^p$ parameters  for the nl $(\nu=2)$ and nnl $(\nu=4)$ SO(3)-PWs. To ensure a simple functional form of the GPDs, we require for the intercepts  $\alpha_\nu^p \le \alpha^p$  for
$\nu\in\{2,4\}$ by employing a half Gaussian distribution. For all other parameters, we employ Gaussian priors to specify the parameter range as obtained from DVCS estimates. Proceeding in this way, we might have over parameterized the GPDs for an estimate at NLO accuracy, however, our objective is here to test if it
is possible to disentangle the skewness- and $t$-dependence.

\begin{figure}[t]
\centering
\includegraphics[width=0.8\textwidth]{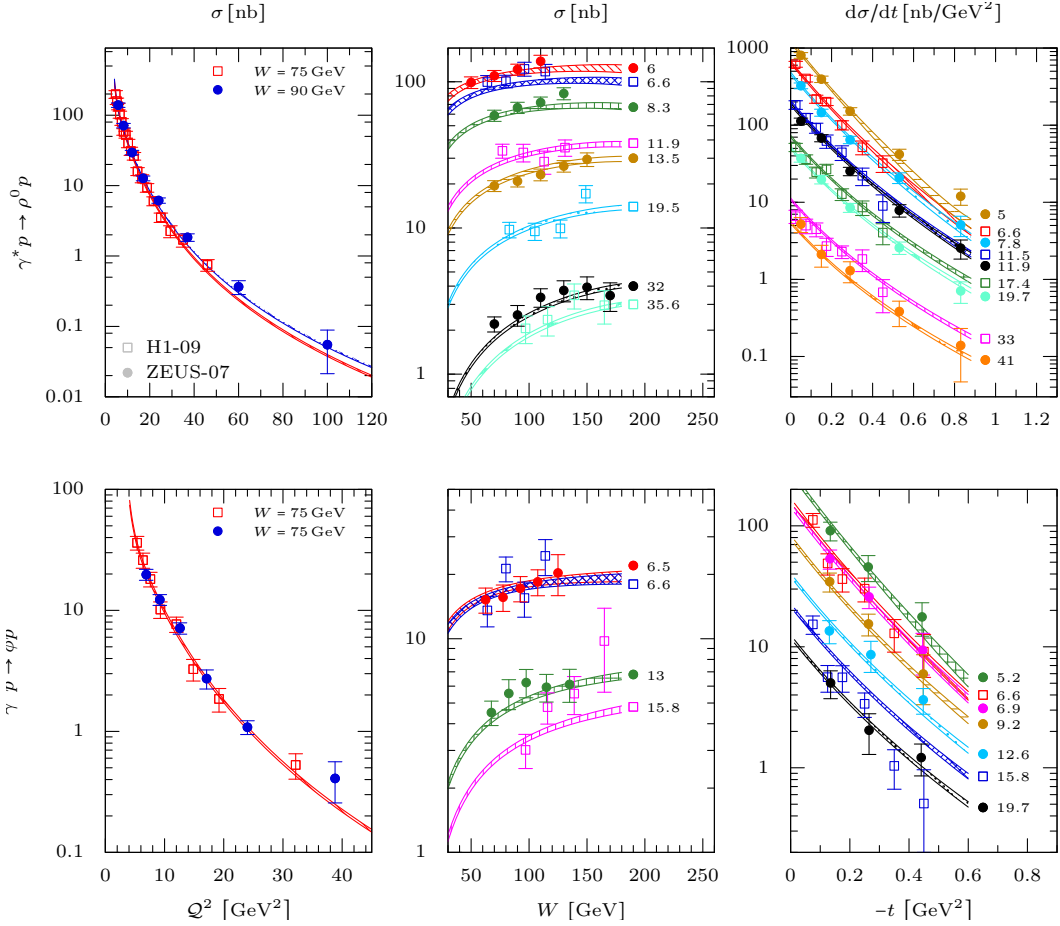}
\caption{
\small
Total cross section vs.~$\Q^2$ (left) and $W$ (middle)  as well as differential one vs.~$-t$ (right) of DV$\!\rho^0$P
(top)  and DV$\phi$P (bottom)
measurements from H1 \cite{Aaron:2009xp} and ZEUS  \cite{Chekanov:2005cqa,Chekanov:2007zr}.
Curves arise from a simultaneous GPD estimate at NLO, where numbers next to them are the  $\Q^2$ values in units of $\unit{GeV^2}$.
}
\label{fig:fit NLO}
\label{fig:DVMP Rho0 NLO}
\label{fig:DVMP Phi NLO}
\end{figure}
We explored the statistical estimate of GPD parameters at LO and NLO accuracy. A combined DIS and DVCS estimate shows that the additional $\alpha_\nu^p$ and $M_\nu^p$  parameters for the higher $\mathrm{SO}(3)$-PWs are not needed to describe the data and we can safely equate them with the parameters for the l-PW, which reproduces the findings in \cite{Kumericki:2009uq}. Including DV$V^0$P data,
where we used asymptotic DAs ($k=0$), we are able to constrain these additional parameters for the higher $\mathrm{SO}(3)$-PWs.  This even allows us to obtain a reasonable description at LO. However, we emphasize that a model comparison of LO and NLO, using the odds ratio \cite{Jaynes2003}, highly favors a NLO description.
In Fig.~\ref{fig:fit NLO} we display our simultaneous NLO estimate of DV$\!\rho^0$P (top) and DV$\!\phi$P (bottom)
measurements versus $\Q^2$ (left), $W$ (middle), and $-t$ (right). We add that as previously the description of DVCS and the selected
DIS data is unproblematic and of the same quality as in \cite{Kumericki:2009uq}. The resulting PDFs can be considered as compatible with standard ones, see e.g., \cite{Kumericki:2009uq}.

\begin{figure}[t]
\centering
\includegraphics[width=16cm]{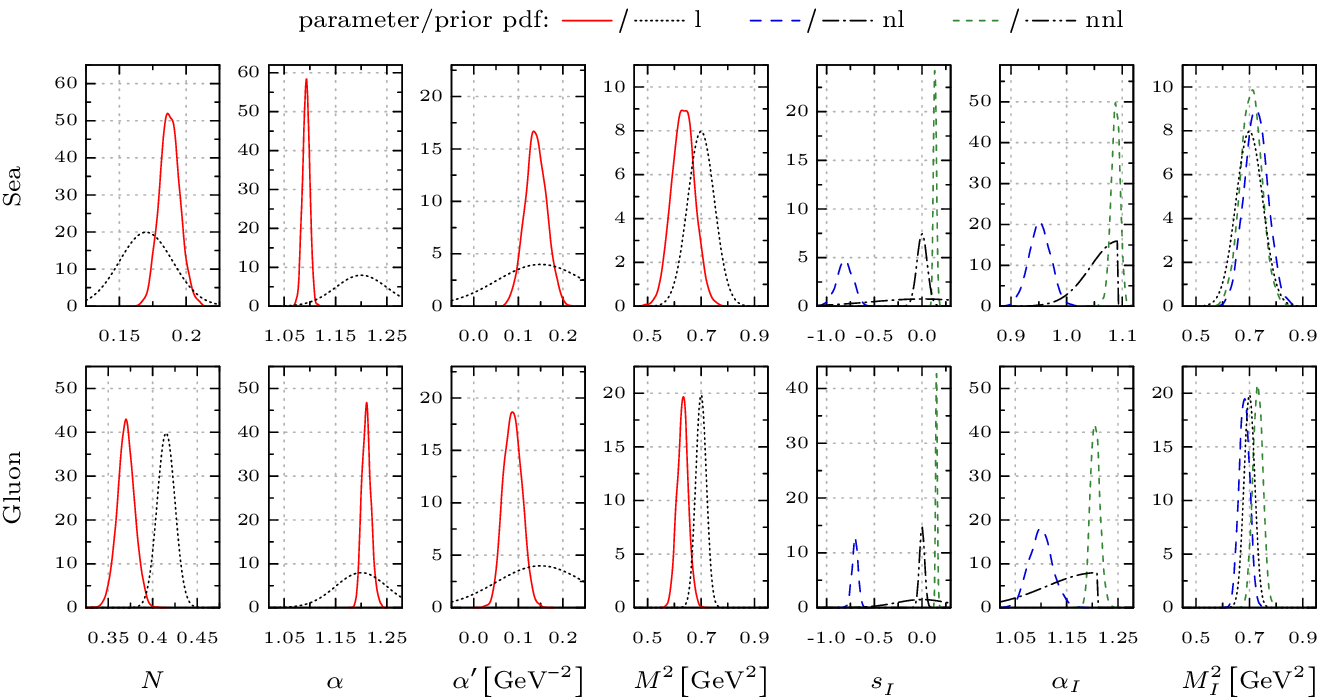}
\vspace{-5mm}
\caption{\small Posterior  (solid, dashed, short dashed) and prior (dotted,dash-dotted,dash-dot-dotted) pdfs of the  parameters (\ref{ps}) [nl-SO(3) PW (dashed, dash-dotted), nnl-SO(3) PW (short dashed, dash-dot-dotted)] for the sea quark (top) and gluon (bottom) GPDs $H^{\rm sea}$ and $H^{\rm G}$, respectively, see (\ref{H_j}) and (\ref{H^p_n}).}
  \label{fig:NLO-pdfs}
\end{figure}
Let us give details of our parameter estimate at NLO accuracy by using asymptotic DAs.
In Fig.~\ref{fig:NLO-pdfs} we show the pdfs for various parameters (\ref{ps}) in comparison to the prior pdfs.
The difference of the posterior pdf (solid) from the prior distribution (dashed) shows the degree by which the data constrain the estimate. If the data does not hold information about a certain parameter, the posterior pdf will be equal to the prior one. In case the information in the data is exactly the prior distribution, we would still obtain a narrower posterior pdf by the rules of error propagation. All parameters besides the cut-off masses are significantly different from the prior distribution. The Regge intercepts for the sea and gluon GPD are similar to the values obtained in \cite{Kumericki:2009uq}, especially the gluon does not become soft and we are in agreement to deep inelastic scattering. We recall that the normalization of the gluon GPD is not strictly fixed by $N^{\rm G}=1-N^{\rm sea}-N^{\rm val}$ rather we allow for a slight variation by employing a corresponding prior.
We find only a slight deviation of the additional parameters from their l-PW values while more flexible DAs, e.g., taking the first three conformal PWs into account, allow us to equate the additional parameters again to the l-PW values. Note that a good LO description can not be reached in this
way.

\begin{figure}[h]
  \centering
\includegraphics[width=14cm]{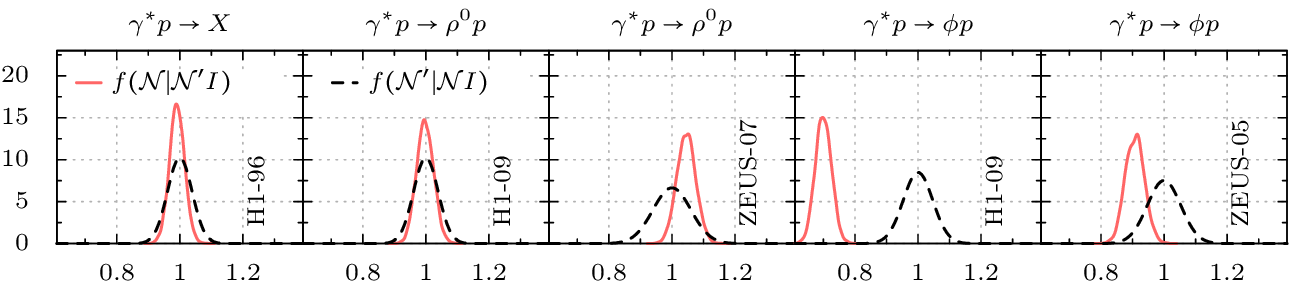}
\vspace{-3mm}
  \caption{\small Posterior pdfs $\cprob[{\cal N}]{{\cal N}^\prime I}$ (solid) of the normalization parameters ${\cal N}_{data}$ and the experimental distributions $\cprob[{\cal N}^\prime]{{\cal N} I}$ (dashed), where ${\cal N}^\prime$ denotes the experimental normalization.}
  \label{fig:NLO-N}
\end{figure}
The pdfs for the normalization parameters ${\cal N}_{\rm data}$ are shown in Fig.~\ref{fig:NLO-N}. As the reader realizes, the normalization  ${\cal N}_{\rm DVMP}$ of DV$\!\rho^0$P and DV$\phi$P from the ZEUS experiment is slightly higher than for the H1 measurements.  This difference is not biased by our assumptions, whereas the low mean value of the DV$\phi$P normalization could be compensated by the shape of the DAs for $\rho^0$ and $\phi$  mesons and/or a different sea quark flavor decomposition.
\begin{figure}[t]
  \centering
\includegraphics[width=14cm]{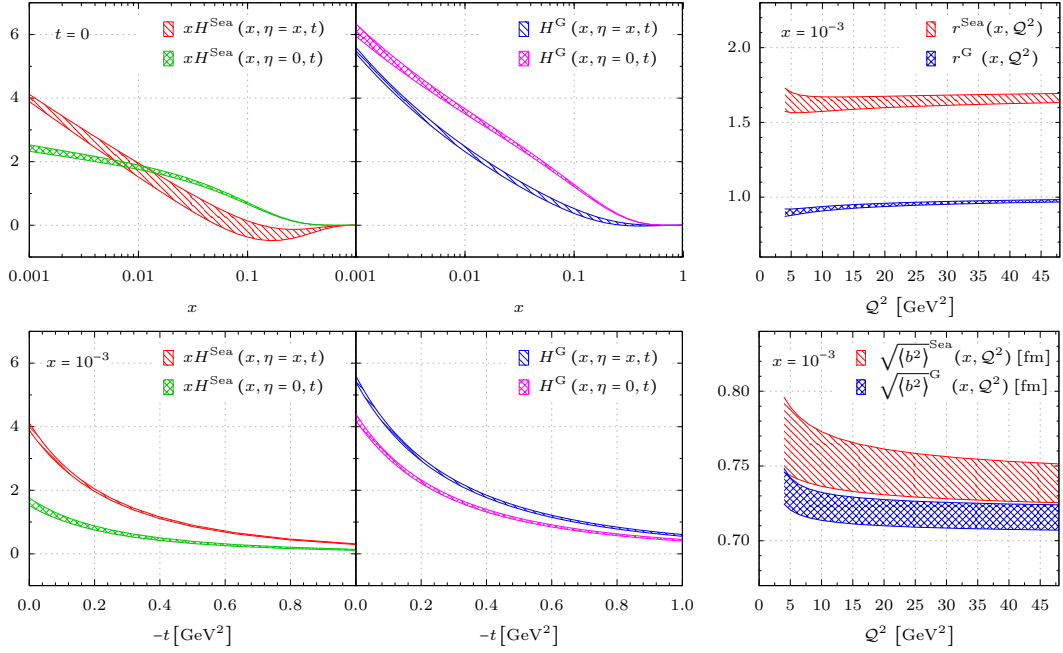}
\caption{\small Left panel: NLO quark (left) and gluon (right) GPDs at the input scale $\Q_0$ on the crossover line and for zero skewness
vs.~$x$ for $t=0$ [upper row] and vs.~$-t$ for $x=10^{-3}$ [lower row]. Right panels: The skewness ratio (\ref{r(x,Q^2)})  [upper row]  and the transverse average (\ref{b^2(x,Q^2)}) [lower row] vs.~$\Q^2$.
}
  \label{fig:PDFGPD_NLO}
\end{figure}


\section{Partonic interpretation}

In Fig.~\ref{fig:PDFGPD_NLO} we show our sea quark and gluon PDFs and GPDs on the cross-over line $\eta=x$, extracted at NLO, as function of $x$ at $t=0$ [top] and of $-t$ at $x=10^{-3}\, (\xB=2\times10^{-3})$ [bottom]. The extrapolation in the $x> 0.016$ region is not constrained by the data. It depends on the $\beta$ values and the model induced skewness effect. We also show the skewness ratio, i.e., the ratio of the GPD on the cross-over line at $t=0$ to the PDF,
\begin{eqnarray}
\label{r(x,Q^2)}
r(x,\Q^2) = \frac{ H (x,\eta=x,t=0,\Q^2)}{ H(x,\eta=0,t=0,\Q^2) }\,,
\end{eqnarray}
which encodes one facet of the accessible GPD information \cite{Kumericki:2009uq}.

As in a DVCS analysis \cite{Kumericki:2009uq} it turns out that the LO findings, i.e., for sea quarks $r^{\rm sea}\approx 1$, increase at NLO to the value $\unit{1.647\pm 0.042}$ while the gluonic one, given at LO by $r^{\rm G} \sim 0.5$, increases to $\unit{0.955\pm 0.025}$. Both of these ratios rise slightly with growing $\Q^2$. This behavior illustrates that GPD and PDF evolution predict a different evolution of the `residue' while the `effective' pomeron intercept evolve in the same manner. Note that the found ratios at NLO are compatible with the LO claim \cite{Shuvaev:1999ce,Martin2009zzb} and tell us that the `pomeron' pole can be roughly put in the complex conformal spin plane, for details see \cite{Kumericki:2009ji}.
Of course, this can not be taken literally, e.g., for gluons we do not have  a ratio that is slightly larger as one as `predicted'. Moreover, at NLO the `predicted' sea quark and gluon ratio will be modified to some extend \cite{Kumericki:2007sa}.

As alluded to in paragraph {\it i}, one important phenomenological goal is  access to the transverse distribution of partons. Already from present small-$\xB$
data this task can be achieved to some certain extend. The simplest example is that the averaged square transverse distance of partons from the proton center can be obtained from the slope measurement at $t=0$,
\begin{eqnarray}
\label{b^2(x,Q^2)}
  \langle\vec b^2 \rangle(x,\Q^2)  = \left.  4\frac{d}{dt} \ln H(x,\eta=0,t,\Q^2) \right|_{t=0}\,,
\end{eqnarray}
which requires both a model dependent extrapolation to $t=0$ and to zero skewness $\eta=0$. In the lower right panel of Fig.~\ref{fig:PDFGPD_NLO} we show the
transverse width $\sqrt{\langle\vec b^2 \rangle}$ at $x=10^{-3}$ as function of $\Q^2$, where we propagated the experimental errors. Our NLO findings, based on a residual dipole ansatz,
for sea quark $\unit[0.745\pm0.017]{fm}$ and gluons $\unit[0.719\pm 0.010]{fm}$ are now a bit smaller as those obtained from a pure DVCS analysis at LO \cite{Kumericki:2009uq}. The errors are the square roots of the variances. Within the uncertainties, the quark and gluon transverse widths coincide. Note that an exponential residual $t$-dependence yields naturally smaller values. Evolution implies that the transverse width slightly decrease with growing $\Q^2$, see discussion in \cite{Kumericki:2009uq}. Moreover, we can now also evaluate the transverse profile density
\begin{equation}
\label{rho(b,x,Q^2)}
  \rho(b,x,\Q^2)
  = \frac{
    \int_{-\infty}^{\infty}\!d^2 \vec\Delta \; e^{i\vec\Delta\vec b} H( x,\eta=0,t=-\vec\Delta^2,\Q^2)
  }{
    \int_{-\infty}^{\infty}\! d^2 \vec\Delta \; H( x,\eta=0,t=-\vec\Delta^2,\Q^2 )
  }\,.
\end{equation}
In this ratio, displayed in Fig.~\ref{fig:rho(b,x,Q^2)},  the functional shape dependence of the residual $t$-dependence cancels mostly,
where, however, the model dependent extrapolation in the large-$t$  and small-$t$ region influences the shape of the profile function at
small and large $|\vec{b}|$, respectively. A closer look at these systematic uncertainties can be found in \cite{AscDieFazINT2011}, while an illustration
of partonic imaging with high precision pseudo data for the proposed eRHIC accelerator is presented  in \cite{Aschenauer:2013qpa}.
\begin{figure}[t]
  \centering
\includegraphics{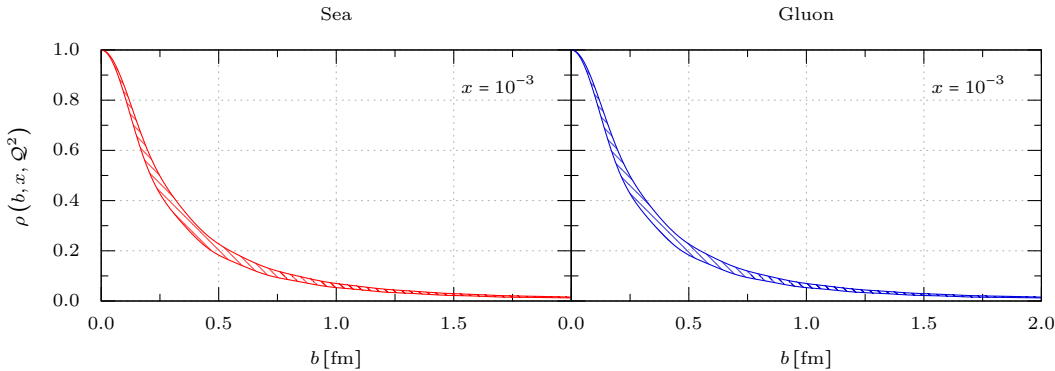}
\vspace{-3mm}
  \caption{\small
  Transverse profile density (\ref{rho(b,x,Q^2)}) for sea quarks (left) and gluons (right) at $x=10^{-3}$ and $\Q^2 = 4\,\GeV^2$
  versus the impact parameter $b=|\vec{b}|$.}
  \label{fig:rho(b,x,Q^2)}
\end{figure}


\section{Conclusions}

Based on the NLO perturbative framework we have performed a global GPD analysis of DIS, DVCS, DV$\rho^0$P, DV$\phi$P of the H1 and ZEUS collaboration. We obtained  very satisfactory description of the data which is a success, claimed more or less impossible a few years ago. To reach this goal we had to make several assumptions.  We neglected the GPD $E$ contribution, employed a standard SU(3) flavor scenario and used a common DA shape. If these assumptions are relaxed, we do not expect from an analysis of presently available data that one can pin down the GPD and DA models  in an unique manner. In particular, an access to the GPD $E$ requires new experiments in the small-$\xB$ region and with a transversally polarized proton beam as it would be possible at eRHIC \cite{Deshpande:2012bu}.

The correlation of DA shape and GPD shape, in particular for the small-$\xB$ region the correlation between the quark sea flavor decomposition and the SU(3) flavor symmetry breaking of DAs, is also present in global GPD estimates, thought on a reduced level. Hence, it would be much desired to have reliable theoretically information on the vector meson DAs. These can come from lattice simulations \cite{Braun:2007zr,Arthur:2010xf}, QCD sum-rules, or the truncated Dyson-Schwinger equation approach (for references see \cite{Segovia:2013eca}). Our GPD results at NLO are compatible with previous ones from a DVCS analysis w.r.t.~both the skewness effect and the transverse distribution of sea quarks and gluons.

We estimate the parameters of our model using Bayesian inference. This gives us several advantages over the least square procedure. Amongst others, we have access to the normalized posterior pdfs for the parameters and all deduced quantities such as GPDs or observables. The availability of the normalization ensures the correct propagation of experimental errors. In addition, the comparison of the posterior pdf to the prior distribution for a parameter shows its sensitivity to the data. Here we employed the method to address the problem whether the $t$-dependence of the three effective SO(3)-PWs differs. This is rather important if one likes to give in future a probabilistic interpretation of the measured $t$-dependence of high precision data. In our analysis of present data we found that it is rather likely that in NLO the $t$-dependence difference is rather mild in the small-$\xB$ region. Our findings encourage us also to include fixed target data in a global GPD analysis which requires updating our GPD models and the DVCS analysis, which is partially based on a dispersive framework.

Let us finally remind that GPDs are process independent quantities, however, they depend on the order in which one is working and on the factorization scheme.  Various variations of the GPD framework were suggested to improve the theoretical control at LO,  which will also implicitly alter the GPD definition. Our results show that this is unnecessary and we argue strongly that if one likes to address GPDs on a quantitative level one should stay within the collinear framework. Surely, it would be highly desirable to reach a deeper theoretical understanding of the processes in question. The most valuable source of information may arise from perturbative calculations that go beyond the present state of art, essentially set up one decade ago, and to address the problem of kinematical higher-twist contributions as done recently for DVCS \cite{Braun:2011dg}.
\\


\section{Acknowledgements}

We are indebted to K.~Passek-Kumeri{\v c}ki for numerous discussions and to K.~Kumeri{\v c}ki for numerical implications of DVMP
formulae, which were used by us for cross checks. We like to thank M.~Diehl, X.~Ji, M.~Strikman and  Ch.~Weiss for valuable discussions.
This work was supported in part by the Joint Research Activity
\emph{Study of Strongly Interacting Matter} (acronym HadronPhysics3,
Grant Agreement No.~283286) under the Seventh Framework Program of the European Community and by the German Ministry of Science and Education (BMBF grant OR 06RY9191 nd 05P12WRFTE).

\small

\end{document}